\documentclass[10pt,nofootinbib, aps, twocolumn, superscriptaddress, groupedaddress,preprintnumbers,prd]{revtex4-2}

\usepackage{bm,bbm}
\usepackage{amssymb}
\usepackage{amsmath}
\usepackage{mathtools}
\usepackage{microtype}

\usepackage{array}
\usepackage{enumerate}

\usepackage{graphicx}
\usepackage{xcolor}
\usepackage{color} 		
\definecolor{darkblue}{rgb}{0.0,0.0,0.5} 		

\usepackage[colorlinks]{hyperref}
\hypersetup{pdftitle={Moduli-dependent KK towers and the swampland distance conjecture on the quintic Calabi-Yau manifold},
	pdfauthor={Anthony Ashmore and Fabian Ruehle},
	linkcolor=darkblue, urlcolor=darkblue, citecolor=darkblue, filecolor=black, linktoc=page}

\usepackage{enumitem}

\let\originalleft\left
\let\originalright\right
\renewcommand{\left}{\mathopen{}\mathclose\bgroup\originalleft}
\renewcommand{\right}{\aftergroup\egroup\originalright}

\newcommand{\im}{\operatorname{Im}}

\begin{document}

\preprint{CERN-TH-2021-032}

\title{Moduli-dependent KK towers and the swampland distance conjecture\\on the quintic Calabi-Yau manifold}
\author{Anthony Ashmore}
\email[]{ashmore@uchicago.edu}
\affiliation{Enrico Fermi Institute \& Kadanoff Center for Theoretical Physics, University of Chicago, Chicago, IL 60637, USA}
\affiliation{Sorbonne Universit\'e, CNRS, Laboratoire de Physique Th\'eorique et Hautes Energies, F-75005 Paris, France}
\author{Fabian Ruehle}
\email[]{fabian.ruehle@cern.ch}
\affiliation{CERN Theory Department, 1 Esplanade des Particules, CH-1211 Geneva, Switzerland}
\affiliation{Rudolf Peierls Centre for Theoretical Physics, University of Oxford, Parks Road, Oxford OX1 3PU, UK}

\begin{abstract}
\noindent We use numerical methods to obtain moduli-dependent Calabi--Yau metrics and from them the moduli-dependent massive tower of Kaluza--Klein states for the one-parameter family of quintic Calabi--Yau manifolds. We then compute geodesic distances in their K\"ahler and complex structure moduli space using exact expressions from mirror symmetry, approximate expressions, and numerical methods and compare the results. Finally, we fit the moduli-dependence of the massive spectrum to the geodesic distance to obtain the rate at which states become exponentially light. The result is indeed of order one, as suggested by the swampland distance conjecture. We also observe level-crossing in the eigenvalue spectrum and find that states in small irreducible representations of the symmetry group tend to become lighter than states in larger irreducible representations.
\end{abstract}

\maketitle

\tableofcontents

\section{Introduction}
\label{sec:Introduction}
In recent years, a wealth of swampland conjectures has been put forward (see~\cite{Palti:2019pca} for a review). They postulate properties that either necessarily arise or never arise in string theory or sometimes more generally in any consistent quantum theory of gravity. A very intriguing conjecture is the (finite version of the) swampland distance conjecture (SDC)~\cite{Ooguri:2006in}, which proposes a relation between the mass of a tower of states and the position in moduli space of the string theory compactification in question. More precisely, compare two theories at points $p_0$ and $p_1$ in the moduli space of the theory. The conjecture then postulates that the theory at $p_1$, a geodesic distance $d(p_0,p_1)$ from $p_0$, has an infinite tower of light particles starting with mass of the order of $e^{-\alpha d(p_0,p_1)}$ for some $\alpha > 0$. The constant $\alpha$, which governs the rate at which the tower of states becomes light and is expected to be $\mathcal{O}(1)$, is important when one wants to study phenomenological implications of the SDC.

Almost simultaneously with the rekindled interest in the swampland program, machine learning techniques were introduced to string theory~\cite{He:2017aed,Ruehle:2017mzq,Krefl:2017yox,Carifio:2017bov} (see~\cite{Ruehle:2020jrk} for a review). While numerical algorithms for computing CY metrics have been studied before~\cite{Donaldson:2005mat,Douglas:2006rr,Braun:2007sn,Headrick:2009jz}, the advent of faster optimizers and computers have made the analysis amendable to machine learning CY metrics~\cite{Ashmore:2019wzb,Anderson:2020hux,Douglas:2020hpv,Jejjala:2020wcc}, even at many different points in complex structure moduli space. Once the (moduli-dependent) Calabi--Yau metric is known, the spectrum of massive string excitations can be computed numerically~\cite{Braun:2008jp,Ashmore:2020ujw}. By varying the moduli, we can then explicitly trace the spectrum of massive string excitations as a function of the position in moduli space. In particular, this will allow us to compute the coefficient $\alpha$ in the SDC \textit{from first principles}.

So far, fast code has mostly been developed for one-parameter CY manifolds. In this paper, we will therefore be studying the one-parameter family of quintics in $\mathbbm{P}^4$ given by the vanishing locus of
\begin{align}
\label{eq:quintic}
z_0^5+z_1^5+z_2^5+z_3^5+z_4^5-5\psi z_0 z_1 z_2 z_3 z_4\,,
\end{align}
where $[z_0:z_1:z_2:z_3:z_4]$ are the homogeneous coordinates. For such families, the rate $\alpha$ has been estimated in the case where a Kaluza--Klein (KK) tower of states becomes light from the fact that the mass of the KK tower is expected to go like
\begin{align}
\label{eq:KKScale}
m_\text{KK}(p_1)\sim \frac{M_\text{Pl}}{r^2}\sim m_\text{KK}(p_0) e^{-\alpha d(p_0,p_1)}
\end{align}
with $r^3 \sim \text{Vol}(X)$ and $\alpha=4/\sqrt3$ in the large radius limit~\cite{Blumenhagen:2018nts,Joshi:2019nzi}.\footnote{In fact,~\cite{Blumenhagen:2018nts} asserts $m_\text{KK}\sim1/r^2$ and~\cite{Joshi:2019nzi} asserts $m_\text{KK}\sim1/r^{1/2}$, leading to different factors of 2 for $\alpha$. We discuss this further in Appendix \ref{app:Einstein}.} We can compare our explicit results to this. However, we do set up the problem such that it generalizes to more complex situations in which the Picard--Fuchs system does not need to be solved analytically, or in which the geodesic trajectory through moduli space is more complicated and thus makes it hard to determine the KK spectrum.

We also note that according to Weyl's law, the eigenvalues $\lambda_n$ of the Laplacian on a real $d$-dimensional Riemannian manifold $X$ with volume $V$ satisfy
\begin{align}
\lim_{n\to\infty} \frac{\lambda_n^{d/2}}{n}\to\frac{(4\pi)^{d/2}\Gamma(1+d/2)}{V}\,,
\end{align}
and hence the eigenvalues go to zero as $m_{\text{KK}}\sim\lambda^{1/2}\sim V^{-1/6}$, i.e.\ the entire KK tower becomes massless. Including the $1/V$ factor of the 4D metric in Einstein frame, we recover the scaling in~\eqref{eq:KKScale}.

This project requires carrying out the following steps:
\begin{enumerate}[noitemsep]
\item Compute the moduli space metric (using either analytic~\cite{Candelas:1990rm} or numeric~\cite{Keller:2009vj} techniques)\label{enum:step1}
\item Compute the geodesics and the geodesic distances in moduli space\label{enum:step2}
\item Compute the CY metric along the moduli space geodesics\label{enum:step3}
\item Compute the massive spectrum from the CY metric\label{enum:step4}
\item Fit a function to the masses and compare with the prediction from the SDC\label{enum:step5}
\end{enumerate}
We describe steps~\ref{enum:step1} to~\ref{enum:step3} in Section~\ref{sec:Geodesics}, steps~\ref{enum:step4} and~\ref{enum:step5} in Section~\ref{sec:Spectrum}, and conclude in Section~\ref{sec:Conclusions}. We discuss the transformation of the metric to Einstein frame in Appendix~\ref{app:Einstein}, and explain how to compute the irreducible representations of the symmetry groups that lead to the degeneracies of the Laplace operator in Appendix~\ref{app:Symmetries}.

\section{Geodesics in moduli space}
\label{sec:Geodesics}
In order to check the SDC, we need to fix two points in the moduli space and then find the shortest geodesic that connects these points. We will therefore need to discuss moduli space geodesics. We will start with a review~\cite{Candelas:1990rm,Blumenhagen:2018nts,Joshi:2019nzi} of geodesics in complex structure moduli space and then briefly comment on the corresponding K\"ahler moduli space results. We will be following~\cite{Candelas:1990rm}.

\subsection{Geodesics in complex structure moduli space}
The K\"ahler potential for the (Weil--Petersson) K\"ahler metric of the complex structure moduli space of a CY manifold $X$ is
\begin{align}
\label{eq:CS-Kahler-Pot}
K_{\text{cs}}=-\ln\left(i\int_X\Omega(\psi)\wedge\bar\Omega(\bar\psi)\right)\,,~~g_{a\bar b}=\partial_a\bar\partial_{\bar b} K_{\text{cs}}\,,
\end{align}
where $\Omega$ is the holomorphic $(3,0)$-form on $X$, $\partial_a=\partial/\partial_{\psi^a}$, $a=1,2,\ldots h^{2,1}(X)$, and $\psi^a$ are the complex structure parameters. The normalization of the K\"ahler potential has been chosen such that, upon dimensional reduction on $X$, the Einstein--Hilbert term is canonically normalized~\cite{Candelas:1989bb,Benmachiche:2008ma}. This ensures that the geodesic distance is given in units of the 4D effective Planck mass.

Choosing a symplectic basis of three-cycles $A^I,B_I\in H_3(X,\mathbbm{Z})$ and dual three-forms $\alpha_I,\beta^I$ with\footnote{Note that there are $2h^{2,1}(X)+2$ three-forms, which can be divided into two pairs of $h^{2,1}(X)+1$ three-forms.} $I=0,1$, normalized such that
\begin{align}
\label{eq:ThreeCycleBasis}
A^I\cap B_J=\int_X \alpha_J\wedge\beta^I=\int_{A^I}\alpha_J=\int_{B_J}\beta^I=\delta^I_J\,,
\end{align}
and all other combinations zero, we can define the period vector
\begin{align}
	\label{eq:periods}
	\Pi=\begin{pmatrix}\mathcal{G}_I\\z^I\end{pmatrix}=\begin{pmatrix}\int_{B_I}\Omega\\\int_{A^I}\Omega\end{pmatrix}\,,
\end{align}
such that
\begin{align}
\Omega\wedge\bar\Omega=z^I\bar{\mathcal{G}}_I-\bar{z}^I\mathcal{G}_I\,.
\end{align}
The periods have been determined analytically in~\cite{Candelas:1990rm} as solutions to a hypergeometric system of Picard--Fuchs equations and can be written in terms of hypergeometric functions.

In~\cite{Keller:2009vj}, a numerical method for computing the moduli space metric has been proposed, which we compare with the exact results. The method proceeds by varying the complex structure, computing a basis of (non-holomorphic) three-forms under the variation, and evaluating the integral appearing in the metric in~\eqref{eq:CS-Kahler-Pot} numerically using Monte Carlo integration. Note that we need to perform the Monte Carlo integral at different points in complex structure moduli space for computing the numerical CY metric anyways. Having obtained the moduli space metric at different points in moduli space, we interpolate the solution and use the interpolated function for further analysis. 

Once we have the periods and the metric, the next step is to compute the Christoffel connection, which, for a K\"ahler metric, is
\begin{align}
\Gamma^c_{ab}=g^{c\bar d}\partial_a g_{b\bar d}\,,\qquad \Gamma^{\bar c}_{\bar a\bar b}=\overline{\Gamma^c_{ab}}\,,
\end{align}
with all other Christoffel symbols zero. We then solve the geodesic equation
\begin{align}
\label{eq:GeodesicsEquation}
\ddot\gamma^c(\tau)+\Gamma^c_{ab}\dot\gamma^a(\tau)\dot\gamma^{b}(\tau) = 0\,,
\end{align}
numerically, where $\gamma^c$ is a curve in complex structure moduli space parameterized by $\tau$ and dots denote derivatives with respect to $\tau$. Finding the geodesic between two points $p_0$ and $p_1$ is then a boundary value problem, which can be solved using a shooting method in Mathematica~\cite{Mathematica}.

Having obtained the geodesic, the geodesic distance along the curve $\gamma$ reads
\begin{align}
\label{eq:GeodesicLength}
d(p_1,p_2)=\int_{\tau_1}^{\tau_2} \text{d}\tau\sqrt{g_{a\bar b}(\gamma(\tau)) \dot\gamma^a(\tau)\dot\gamma^{b}(\tau)}\,,
\end{align}
with $\gamma(\tau_i)=p_i$. Subsequently, the geodesic length is computed using numerical integration via Mathematica's function \texttt{NIntegrate}.

\subsection{Geodesics in K\"ahler moduli space}
For the K\"ahler moduli, one can proceed completely analogously. The K\"ahler moduli K\"ahler potential can be computed from the complex structure moduli K\"ahler potential of the mirror $\tilde{X}$ of $X$. In the large volume regime, the properly normalized K\"ahler moduli K\"ahler potential is
\begin{align}
\label{eq:KahlerModuliKahlerPot}
K_{\text{k}}	=-\ln\left(\int_X \frac{-i}{3!}\mathcal{J}(t)\wedge \mathcal{J}(t)\wedge \mathcal{J}(t) \right)\,,\;g_{i\bar{j}}=\partial_i\bar\partial_{\bar j} K_{\text{k}}\,,
\end{align}
where $\partial_i=\partial/\partial_{t^i}$, $i=1,2,\ldots h^{1,1}(X)$. The $t^i$ are the complexified K\"ahler parameters for the complexified K\"ahler form and are defined as follows: given a basis of two-cycles $\mathcal{C}_i$ and dual $(1,1)$-forms $J^i$ analogous to the three-cycles and three-forms in~\eqref{eq:ThreeCycleBasis}, we can write
\begin{align}
t^i=\int_{C_i}\mathcal{J}=i\int_{C_i} J+\int_{C_i}B\,,
\end{align}
where $J=r^i J_i$ is the (real) K\"ahler form of $X$, and $B$ is the Kalb--Ramond $B$-field. Then
\begin{align}
\int_X \mathcal{J}(t)\wedge \mathcal{J}(t)\wedge \mathcal{J}(t)=\frac{1}{6}d_{ijk}(t^i-\bar{t}^i)(t^j-\bar{t}^j)(t^k-\bar{t}^k)
\end{align}
where
\begin{align}
d_{ijk}=\mathcal{C}_i\cap\mathcal{C}_j\cap\mathcal{C}_k=\int_X J^i\wedge J^j \wedge J^k
\end{align}
are the triple intersection numbers on $X$. If one defines
\begin{align}
r^i\coloneqq\im(t^i)\quad\Rightarrow\quad K_{\text{k}}=-\ln\left(\int_X d_{ijk}r^ir^jr^k\right)\,,
\end{align}
one still needs to take derivatives $\partial/\partial_{t^{i}}$ rather than $\partial/\partial_{r^{i}}$ when computing the metric; otherwise, the result will differ by an overall factor, which is essential for the proper normalization of the Einstein--Hilbert term.

\subsection{Geodesics for the one-parameter quintic}
\subsubsection*{Complex structure moduli space}
For the one-parameter family of quintics~\eqref{eq:quintic}, the periods are functions of $\psi^5$ rather than $\psi$. This can also be seen from the fact that $\psi\to e^{2\pi i/5}\psi$ can be undone by a coordinate redefinition, e.g.\ by sending $z_0\to e^{-2\pi i/5} z_0$ in~\eqref{eq:quintic}. Hence, we need to consider the range 
\begin{align}
\psi=\rho e^{i\varphi}\,,\qquad 0\leq \rho\leq\infty\,,~~0\leq\varphi<2\pi/5
\end{align}
(as we shall see below, an additional $\mathbbm{Z}_2$ symmetry further restricts this range to $\pi/5$). There is a conifold singularity at $\psi=1$, since the hypersurface is not transverse, i.e.\ $p=\partial_{z_\alpha}p=0$ has a solution in $\mathbbm{P}^4$, e.g.\ $z_\alpha=1$ for $\alpha=0,1,\ldots,4$. This singularity is at finite geodesic distance in complex structure moduli space. For $\rho\to\infty$, the quintic is given by the singular hypersurface $\prod_\alpha z_\alpha=0$. This degeneration is at infinite geodesic distance. In order to move $\mathcal{O}(1)$ in Planck units, we have to change $|\psi|$ by several orders of magnitude, and the larger $|\psi|$ is, the larger the change needs to be. Hence, in order to get trans-Planckian field displacements, we will be necessarily moving towards the infinite-distance point.

\begin{figure*}[t]
\includegraphics[height=.24\textwidth]{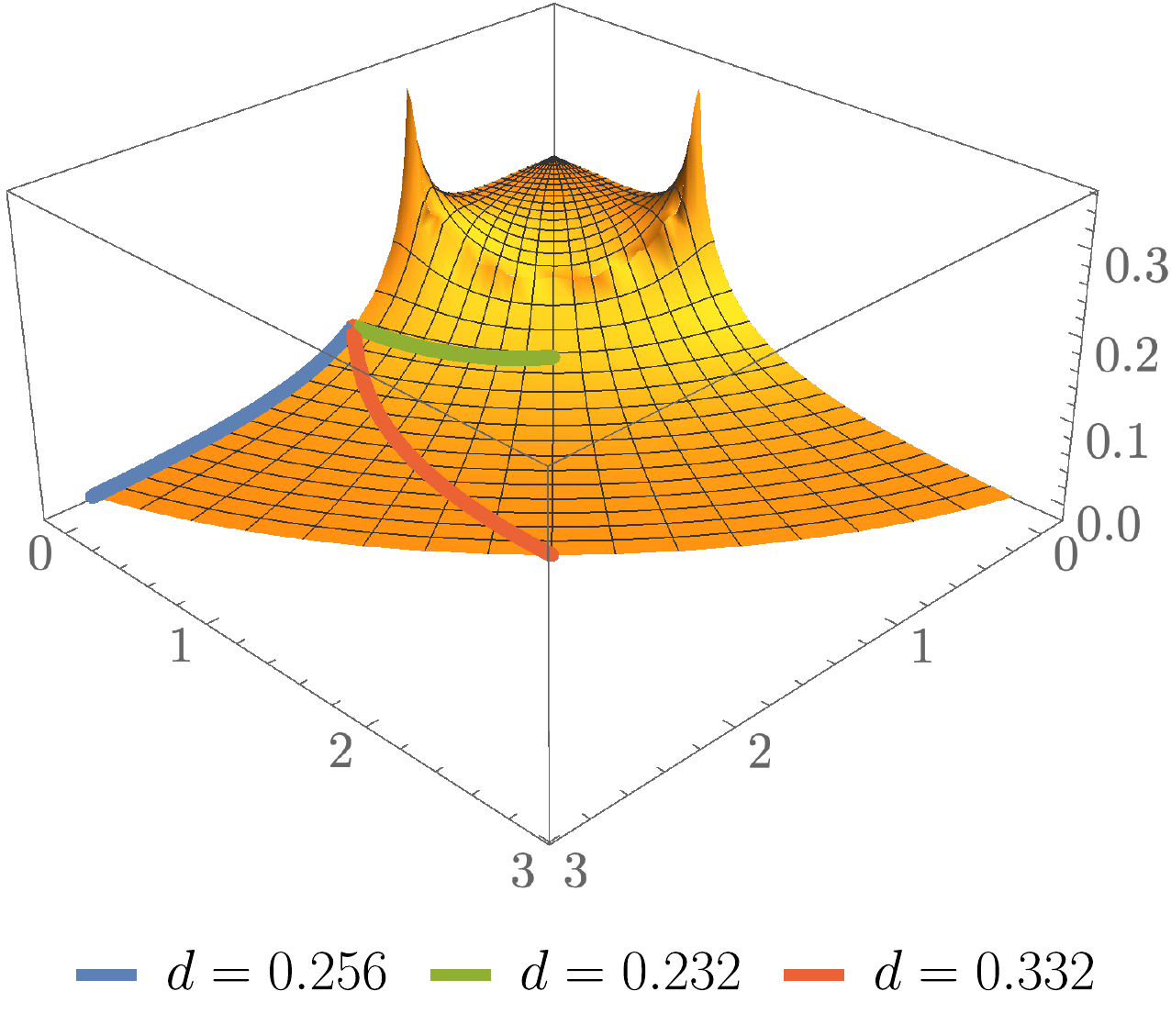}~~
\includegraphics[height=.24\textwidth]{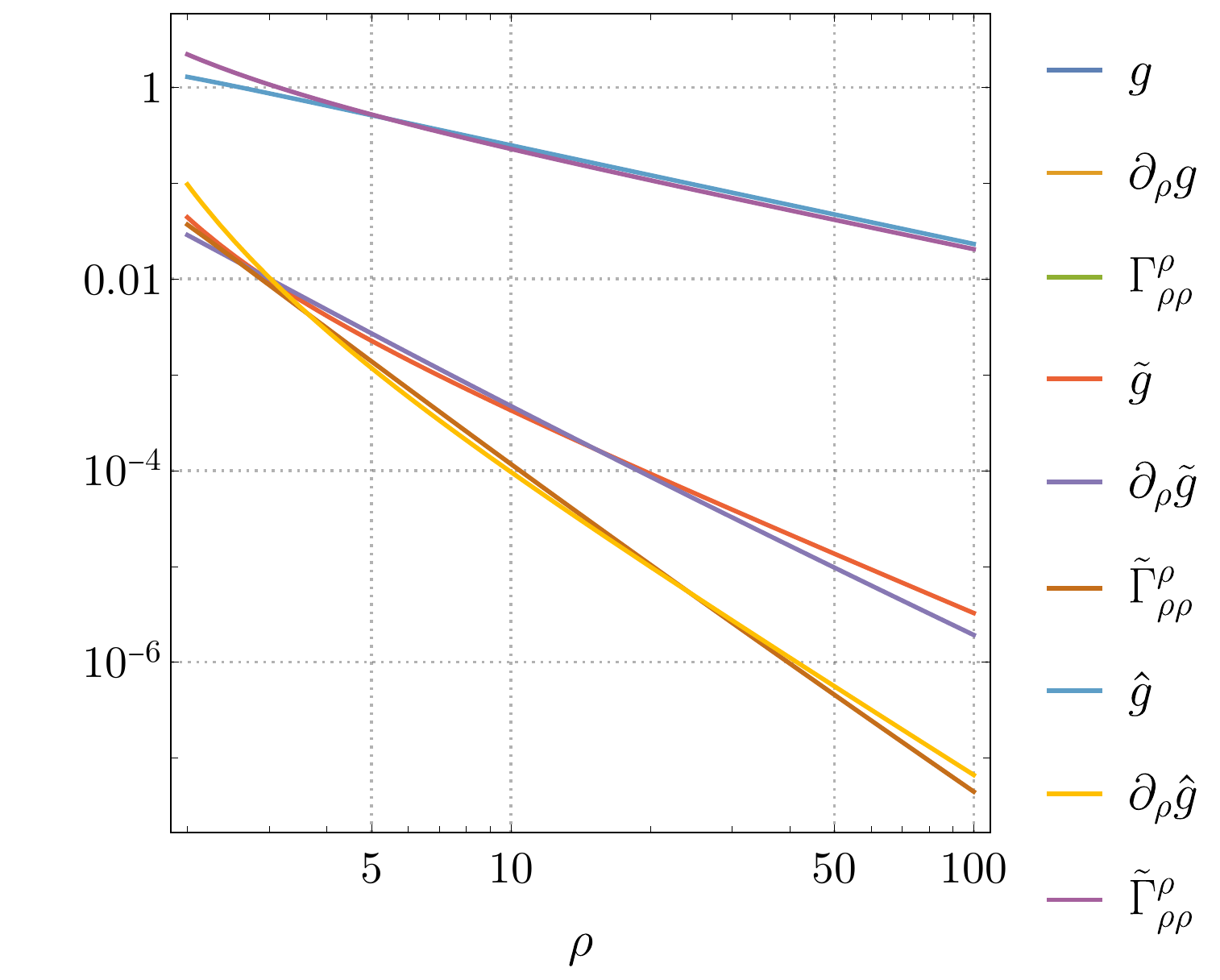}~~
\includegraphics[height=.24\textwidth]{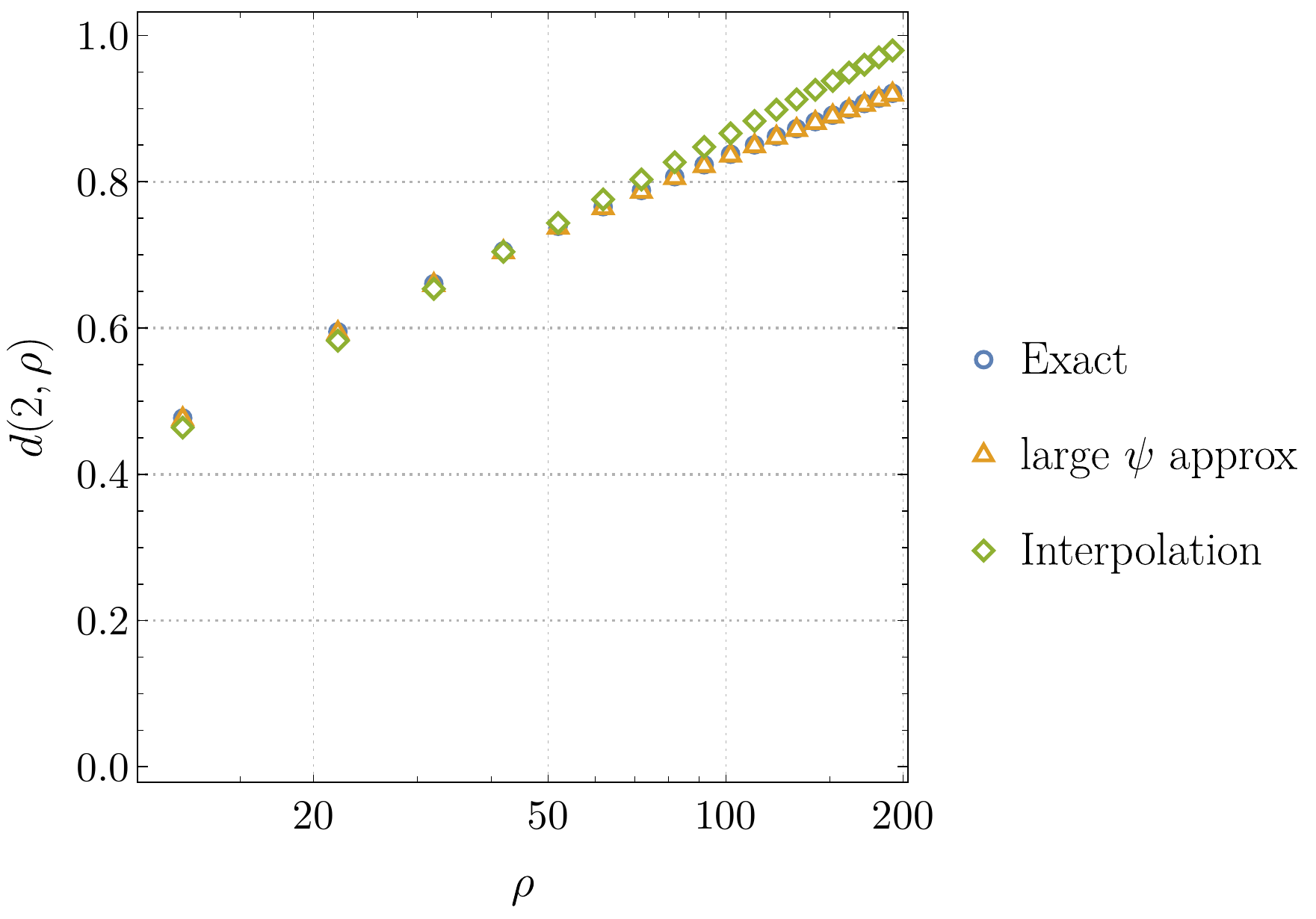}
\caption{(Left): Moduli space metric for $\psi=\rho e^{i\varphi}$ with $0\leq\rho\leq3$, $0\leq\varphi\leq2\pi/5$. (Middle): The quantities $g$, $\partial_\rho g$, $\Gamma^\rho_{\rho\rho}$ for exact metric (no decoration), large $|\psi|$ approximation~\eqref{eq:LargePsiApprox} (tilded), and interpolated from numeric determination of the metric (hatted). (Right:) Geodesic distance $d(2,\psi)$ for $2\leq\psi\leq200$.}
\label{fig:GeodesicsPlots}
\end{figure*}

We plot the moduli space metric in Figure~\ref{fig:GeodesicsPlots} (left) for $0\leq\rho\leq3$ and $0\leq\varphi<2\pi/5$. Note that the $x$- and $y$-axes are to be identified; in particular, there is only one (conifold) singularity at $r=1, \varphi=0\equiv2\pi/5$, and there is a $\mathbbm{Z}_2$ symmetry along the line $\varphi=\pi/5$. Moreover, as one can see, the metric is nearly independent of $\rho$ for $\rho\ll1$, and essentially independent of $\varphi$ for $\rho\gg1$. Indeed, for large $\rho$, the hypergeometric functions describing the periods can be expanded to yield~\cite{Candelas:1990rm}
\begin{align}
\label{eq:LargePsiApprox}
g_{\psi\bar\psi}=\frac{3}{\rho^2\ln^2(5\rho)}\left(1-\frac{48\zeta(3)}{25\ln^3(5\rho)}+\ldots\right)\,.
\end{align}
Hence, the geodesic distance grows slowly with $|\psi|$ for large $|\psi|$.

Let us next discuss the geodesics. We plot three example geodesics onto the metric in Figure~\ref{fig:GeodesicsPlots} (left) for $(p_1,p_2)=(1.4,3)$ (blue), for $(p_1,p_2)=(1.4,1.4\, e^{2\pi i/10})$ (green) and for $(p_1,p_2)=(1.4,3\,e^{2\pi i/10})$ (red). Their geodesic lengths are of the order $0.25$ to $0.3$ and are also given in the figure. Since we want to capture the ``generic'' behavior of the spectrum, we start a bit away from the conifold point. In what follows, we start at $|\psi|=2$.\footnote{For plotting the geodesics, we started at $\psi=1.4$ for better visibility of the geodesics in the plot.} From solving the geodesic equation, one finds that $\varphi$ stays constant when starting and ending at $\varphi=0$; in any case, as discussed above, the $\varphi$-dependence is mild for larger $\psi$. 

For ease of exposition, we set $\varphi=0$ at $p_1$ and $p_2$, such that we can focus on the geodesic $\gamma(\tau)=\rho(\tau)$ and compare three different methods for computing it:
\begin{itemize}
\item The exact Weil--Petersson metric obtained from analytic continuation of hypergeometric functions,
\item The large-$|\psi|$ approximation~\eqref{eq:LargePsiApprox} (indicated by a tilde),
\item The interpolated metric from numerical approximation to the metric (indicated by a hat).
\end{itemize}

For the numerical result, we use the algorithm of~\cite{Keller:2009vj} and compute the moduli space metric numerically at points $\rho\in[2,12,22,\ldots,202]$. We use 30,000 points for the numerical integration. After that, we interpolate the points. Several methods seem to work here. We used our domain knowledge that the underlying Picard--Fuchs system leads to hypergeometric functions and fitted a function
\begin{align}
\hat{g}(\rho)=\frac{a}{\rho^2}+\frac{b}{(\rho\ln\rho)^2}\,,
\end{align}
using the $L_1$ norm (since the values of $g$ get very small, higher norms would essentially only fit the first few data points) and Tikhonov regularization. In cases without prior domain knowledge, Mathematica's~\texttt{FindFormula} can give an idea of which basis functions to use. 

We plot the results we obtain for the metric, the derivative of the metric, and the Christoffel symbol $\Gamma^\rho_{\rho\rho}$ using the three different methods in Figure~\ref{fig:GeodesicsPlots} (middle). As we can see from the plot, the results for the large-$|\psi|$ approximation essentially agrees with the exact result already for $\rho\sim\mathcal{O}(\text{few})$, and the results from interpolating the numerical approximation of the moduli space metric are not too far off. 

The geodesic equation cannot be solved analytically, not even in the large-$|\psi|$ approximation (there, it can be integrated in terms of incomplete Gamma functions, but this cannot be inverted and we refrain from giving the expression). We hence solve the differential equation~\eqref{eq:GeodesicsEquation} numerically using Mathematica's~\texttt{NDSolve} with a shooting method as explained above. Since the start and end points are chosen at $\varphi=0$, the geodesics move radially outward. 

Finally, we compute geodesic distances for 
\begin{align}
p_1=\rho(\tau_1)=2\,,\quad p_2=\rho(\tau_2)\in[2,12,22,\ldots,202]\,.
\end{align}
We compute this distance using a numeric integration algorithm (Mathematica's~\texttt{NIntegrate} with standard parameters) of equation~\eqref{eq:GeodesicLength}. The results for the distances obtained for the exact, approximated, and interpolated metric are plotted in Figure~\ref{fig:GeodesicsPlots} (right). We see that changing $\psi$ by 200 (from $2$ to $202$) corresponds to almost one Planck distance. We also find good agreement between the numerical and the exact methods. 

The numerical results can be further improved by including, for each choice of complex structure,  more points on the CY when performing the Monte Carlo integral. The computation time only grows linearly with the number of points, and finding points on the manifold is very quick, so this is indeed feasible. It will, however, become unfeasible in high-dimensional complex structure moduli spaces, since the number of points at which we need to compute the metric is exponential in the number of dimensions; the same limitation applies for computing the moduli-dependent Calabi--Yau metric as well.

\subsubsection*{K\"ahler moduli space}
Let us also discuss geodesics in K\"ahler moduli space. The mirror map allows us to relate the complex structure modulus $\psi$ with the K\"ahler modulus $t$ of the mirror via
\begin{align}
\label{eq:MirrorMap}
t=\frac{z^1}{z^0}\sim-\frac{5}{2\pi i}\ln(5\psi)~~\Rightarrow~~g_{t\bar t}\sim\frac{3}{4\im(t)^2}\,,
\end{align}
where $z^I$ are the periods~\eqref{eq:periods} and we have used the large $|\psi|$ approximation. The mirror map hence shows that the large complex structure limit is equal to the large volume K\"ahler moduli space metric. Indeed, the triple intersection number of the quintic is $d_{111}=5$, such that we readily find from~\eqref{eq:KahlerModuliKahlerPot}
\begin{align}
K_{\text{k}}=-3\ln(t-\bar t)~~\Rightarrow~~g_{t\bar t}=\frac{3}{4\im(t)^2}\,.
\end{align}

Note that the mirror map~\eqref{eq:MirrorMap} implies that the numerical value of the K\"ahler parameter is changing exponentially with the value of the complex structure parameter in the large parameter limit. We can hence use the simple form of the metric to compute the geodesics at large volume. The geodesic length~\eqref{eq:GeodesicLength} can be computed directly as
\begin{align}
d(p_1,p_2)=\int_{\tau_1}^{\tau_2}\text{d}\tau\frac{\sqrt3}2\frac{\dot r}{r}=\frac{\sqrt3}{2}(\ln p_2-\ln p_1)\,.
\end{align}
This, together with equation~\eqref{eq:KKScale} then implies $\alpha=4/\sqrt{3}$. Expressed in terms of $\psi$, we get the double logarithm
\begin{align}\label{eq:psi}
d(p_1,p_2)=\frac{\sqrt{3}}{2}\ln\left[\ln p_2 - \ln p_1 \right]\,.
\end{align}
Note that hypergeometric Picard--Fuchs systems and the mirror map $t\sim\ln(\psi)$ appear more generally in CY compactifications~\cite{Hosono:1993qy,Blumenhagen:2018nts,Joshi:2019nzi}, and similar considerations apply there.

\section{CY metric and massive spectrum}
\label{sec:Spectrum}

Having obtained the relation between moduli space positions and the geodesic distance traversed, we now need to compute the spectrum of massive modes in the effective theory and observe how it varies with $\psi$. For what follows we will focus on scalar modes in four dimensions, which correspond to eigenfunctions of the scalar Laplace operator on the quintic. Each eigenfunction of this operator with eigenvalue $\lambda_n>0$ gives rise to a massive excitation in four dimensions with mass $m^2_n\sim \lambda_n$. These modes fill out a subset of the KK tower (with the remaining modes coming from $(p,q)$-forms on the quintic, and so on). 

We compute the spectrum in two steps: first we find a numerical approximation of the Ricci-flat metric on the quintic for a fixed value of $\psi$; we then compute matrix elements of the Laplace operator and find the eigenvalues of the resulting matrix. Let us quickly review these ideas before presenting our results.

\subsection{Numerical CY metrics}

There are no known analytic expressions for metrics on non-trivial Calabi--Yau threefolds. Fortunately, there are now many ways to obtain such metrics numerically, including position-space methods~\cite{Headrick:2005ch} and a number of spectral methods such as balanced metrics~\cite{Donaldson1,Donaldson2,Douglas:2006rr,Braun:2007sn}, optimal metrics~\cite{Headrick:2009jz}, and, more recently, machine learning and neural networks~\cite{Anderson:2020hux,Douglas:2020hpv,Jejjala:2020wcc}. 

We use the optimal metrics approach. The basic idea behind this, and the other spectral methods, is to make an ansatz for the K\"ahler potential of the Ricci-flat metric which depends on some constant (but moduli-dependent) parameters. One can then vary these parameters in order to find an approximation to the honest Ricci-flat metric within the space of metrics described by the ansatz. 

One first picks a positive integer $k$ and chooses a basis $\{s_\alpha\}$ for the degree-$k$ monomials of the homogeneous coordinates on $\mathbbm{P}^4$ modulo the equation defining the quintic~\eqref{eq:quintic}. The ansatz for the K\"ahler potential is then given by a generalization of Fubini--Study:
\begin{equation}\label{eq:K_ansatz}
	K = \frac{1}{\pi k} \ln \bigl(s_\alpha h^{\alpha\bar\beta} \bar{s}_{\bar\beta}\bigr)\,,
\end{equation}
where $\alpha,\bar\beta=1,\ldots,\dim \{s_\alpha\}$ and $h^{\alpha\bar\beta}$ is a moduli-dependent (but coordinate-independent) hermitian matrix. The corresponding metric on $X$ is $g_{m \bar n}=\partial_m \bar{\partial}_{\bar n}K$. These so-called ``algebraic metrics'' give a subspace of all possible metrics on $X$. The idea is then to vary the parameters $h^{\alpha\bar\beta}$ to find the algebraic metric closest to the Ricci-flat metric. Note that the integer degree $k$ of the sections controls the size of the basis $\{s_\alpha\}$ and thus the number of parameters in $h^{\alpha\bar\beta}$, and so a larger $k$ allows for a better approximation of the Calabi--Yau metric.

The question then is how to find the ``best'' choice of $h^{\alpha\bar\beta}$. This can be done by noting that on a Calabi--Yau, a K\"ahler metric $g_{m\bar{n}}$ is Ricci-flat if and only if $\eta\coloneqq\Vert \Omega \Vert^2/\det g_{m\bar n}$ is constant. One can then find the optimal metric by varying $h^{\alpha\bar\beta}$ to minimize the error in $\eta = \text{constant}$ integrated over the manifold. The accuracy of the resulting metric is often reported using the ``$\sigma$-measure'', defined in \cite{Douglas:2006rr,Braun:2007sn}.
	
For examples with a large discrete symmetry, such as our one-parameter family of quintics,\footnote{The Fermat quintic ($\psi=0$) admits a $(S_5\times \mathbbm{Z}_2)\ltimes(\mathbbm{Z}_5)^4$ symmetry which is broken to $(S_5\times \mathbbm{Z}_2)\ltimes(\mathbbm{Z}_5)^3$ for $\psi\in\mathbbm{R}^*$, cf.\ Appendix~\ref{app:Symmetries}.\label{symmetry}} the number of independent parameters that one must minimize over is greatly reduced~\cite{Headrick:2009jz}. The resulting optimal metrics are highly accurate and can be computed using the Mathematica package at \cite{headrickmathematicapackage}. We chose to compute the numerical metrics at $k=8$ and used ${10}^4$ points for the minimization procedure. The resulting metric was then evaluated for a further $N_p=2\times{10}^6$ points both to check the accuracy of the approximate metrics and as inputs for the calculation of the spectrum.

The points on $X$ were generated using the intersecting lines importance sampling method from \cite{Douglas:2006rr,Braun:2007sn}. For large values of $\psi$, one finds this greatly undersamples some regions of $X$ and oversamples others, leading to an effective number of points $N_{\text{eff}}$ much smaller than the desired number $N_p$. To ameliorate this, we generate a much larger set $N_s\gg N_{\text{eff}}>N_p$ of points and resample from $N_s$ by drawing elements with a prior given by the weights of the points. This is known as ``sequential Monte Carlo'' or ``sequential importance sampling and resampling'', and leads to a uniform distribution of points on the CY (i.e.\ all points then have weight 1).

\subsection{The spectrum of the Laplace operator}

The eigenfunctions of the Laplace operator $\Delta$ on $X$ are defined by
\begin{equation}\label{eigenvalue_equation}
	\Delta \phi = \lambda \,\phi \,,
\end{equation}
where $\lambda$ is the eigenvalue and $\phi$ the corresponding eigenfunction. The spectrum of $\Delta$ is the set of eigenvalues $\{\lambda_n\}$. The zero-mode ($\lambda_1=0$) of $\Delta$ is unique up to scale and is simply the constant function. Since $X$ is compact, the eigenvalues are discrete and the eigenspaces are finite-dimensional. Furthermore, since the Calabi--Yau metrics for the one-parameter family of quintics admit discrete symmetries, the eigenvalues can appear with multiplicities $\mu_n$ given by the dimensions of the irreducible representations (irreps) of the symmetry group~\cite{Braun:2008jp}. We compute these using GAP as explained in Appendix~\ref{app:Symmetries}; see \eqref{eq:irreps} for the irreps that occur.

Since $\Delta = \delta \text{d}=\star\, \text{d}\star \text{d}$ on scalar functions, the Laplacian depends on the choice of metric on $X$, which is why we need the Calabi--Yau metric to compute the spectrum. Note also that the eigenvalues scale with the volume $V$ (as measured by $g$): since $\Delta\sim g^{m\bar n}$ and $V = \int_X J^3/6\sim\det{g_{m\bar n}}$, we find that $\Delta$ and consequently $\lambda_n \sim V^{-1/3}$. We normalize the volume of $X$ to one in what follows.

The spectrum and eigenfunctions can be computed from the matrix elements of $\Delta$~\cite{Braun:2008jp}. Let $\{\alpha_A\}_\infty$ be some basis for the (infinite-dimensional) space of complex functions on $X$. An eigenfunction of $\Delta$ can be expanded in this basis as $\phi = \phi^A \alpha_A$. The eigenvalue equation \eqref{eigenvalue_equation} can then be written as
\begin{equation}
	\Delta_{AB}\phi^B = \lambda\, O_{AB} \phi^B \,,
\end{equation}
where $\Delta_{AB}=\langle\alpha_A,\Delta \alpha_B\rangle$ and $O_{AB}=\langle\alpha_A,\alpha_B\rangle$, which measures the non-orthonormality of the basis, are computed using the usual inner product on functions and Monte Carlo integration over $X$. This is then a ``generalized eigenvalue problem'' where $\lambda$ gives the eigenvalue of \eqref{eigenvalue_equation} and $\phi^A$ describes the eigenfunction in the chosen basis. 

In practice, one cannot compute with infinite-dimensional matrices, so as in~\cite{Braun:2008jp,Ashmore:2020ujw} we restrict to an approximate finite-dimensional basis, which we denote by $\{\alpha_A\}$. A natural choice are the functions
\begin{equation}
	\{\alpha_A\} = \frac{\bigl\{s^{(k_\phi)}_\alpha \, \bar{s}^{(k_\phi)}_{\bar\beta}\bigr\}}{\bigl(|z_0|^2+\ldots+|z_4|^2\bigr)^{k_\phi}} \,,
\end{equation}
where $\{s^{(k_\phi)}_\alpha\}$ are the degree-$k_\phi$ monomials of the homogeneous coordinates on $\mathbbm{P}^4$. The resulting basis consists of the eigenfunctions for the first $k_\phi + 1$ eigenvalues of the Laplacian on $\mathbbm{P}^4$ restricted to $X$, giving it the interpretation of a spectral expansion~\cite{Headrick:2005ch}. As with the metric, a larger value of $k_\phi$ gives a larger basis which leads to a better approximation of the eigenfunctions and spectrum of $\Delta$. 

For our calculations we chose $k_\phi=3$ which gives $\dim\{\alpha_A\} = 1{,}225$, allowing us to compute the first 1,225 eigenvalues in the spectrum. Note that we restrict our discussion to the first 100 or so eigenvalues since these low-lying modes are well described by the $k_\phi=3$ basis. At the upper end of the spectrum, one expects (and observes) a loss of accuracy due to the finite size of the approximate basis.

\subsection{Numerical results}

\begin{figure*}[t]
	\includegraphics[height=.245\textwidth]{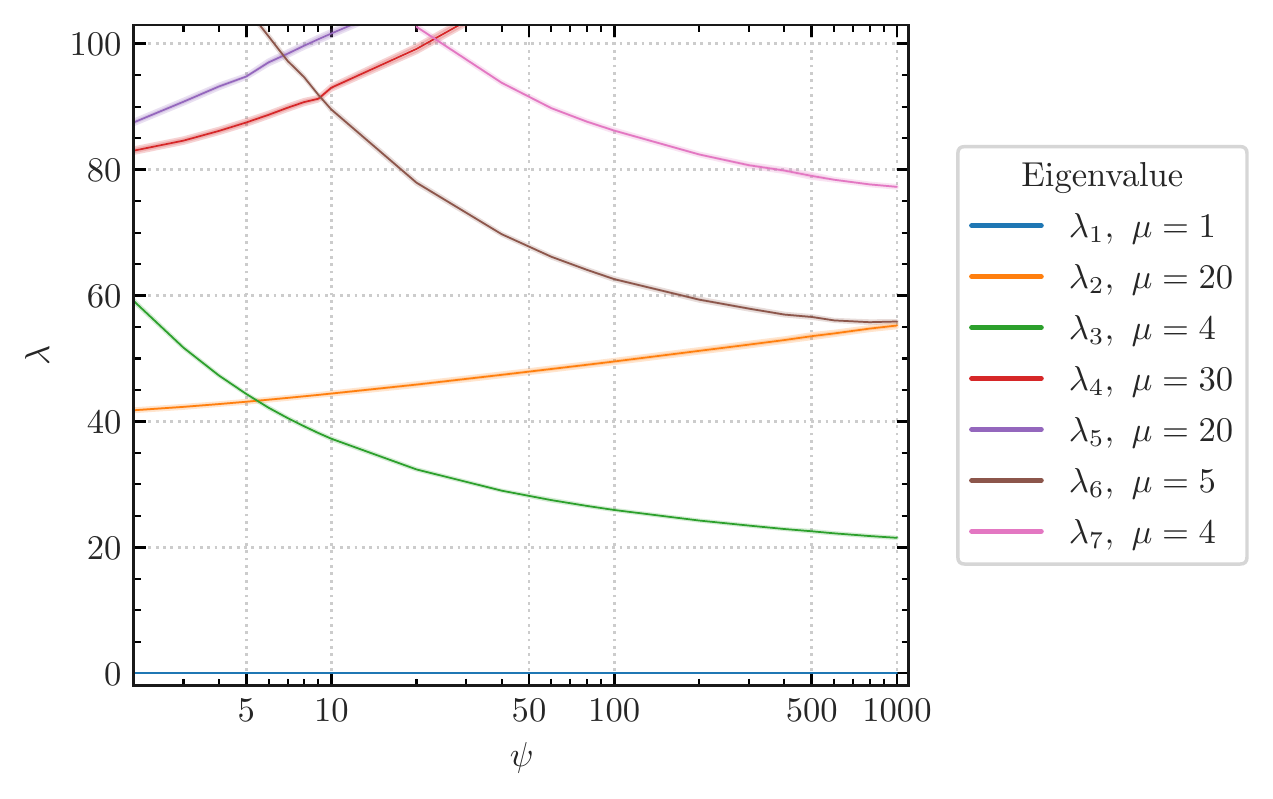}~~~~~~~~
	\includegraphics[height=.24\textwidth]{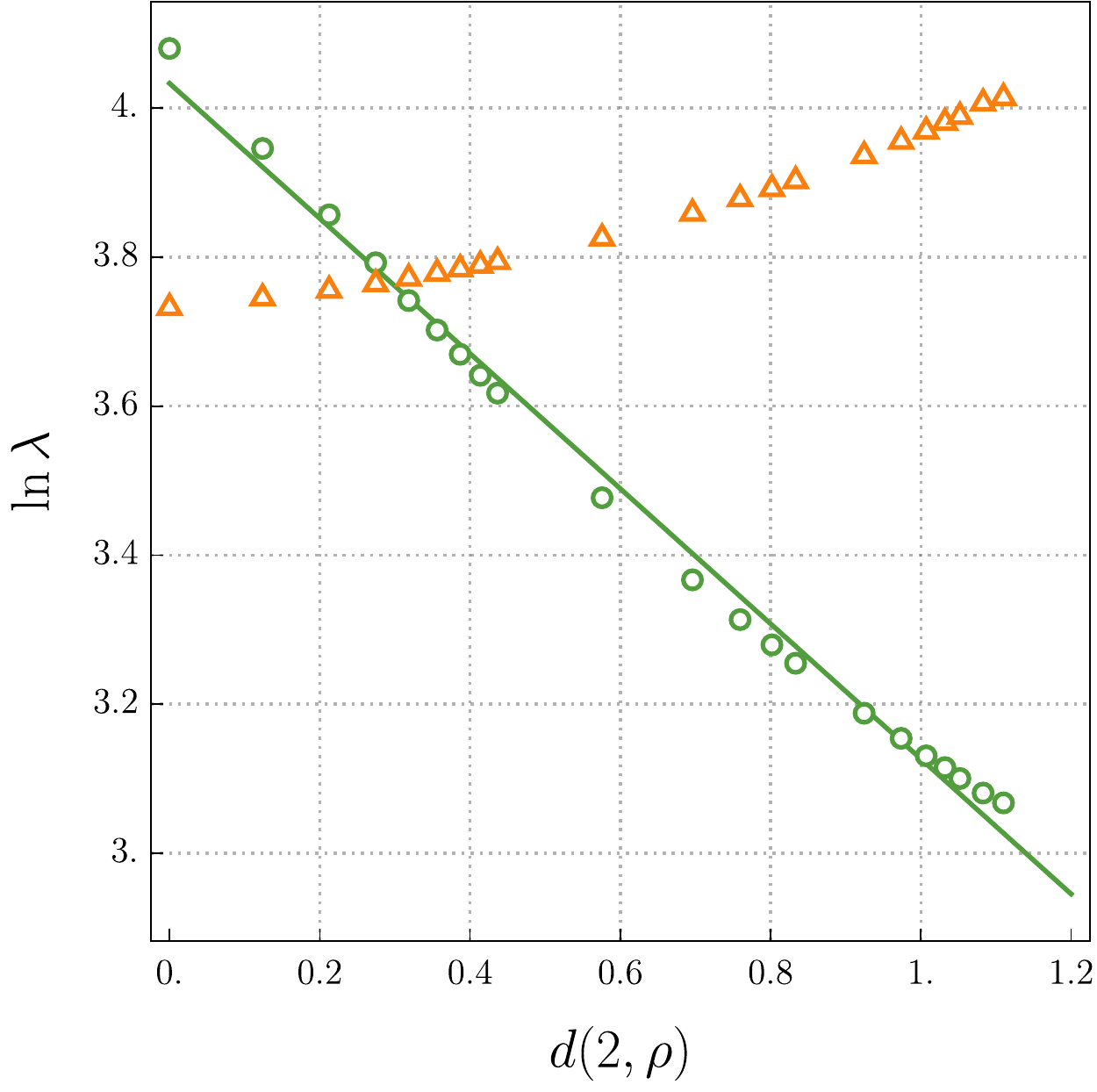}
	\caption{(Left): Low-lying eigenvalues varying with $\psi$ for $2\leq\psi\leq1000$. (Right): First two non-zero eigenvalues of the scalar Laplace operator varying with geodesic distance $d(2,\rho)$ from $\psi=2$. The line of best fit is $56.4 \,e^{-(0.906\pm0.034) \,d(2,\rho)}$.}
	\label{fig:EigPlots}
\end{figure*}

We computed the scalar spectrum of the Laplace operator for the one-parameter family of quintics \eqref{eq:quintic} for $2\leq \psi \leq 1000$. The $\sigma$-measure for the approximate metrics ranged from a minimum of $3\times{10}^{-4}$ for $\psi=2$ to a maximum of $0.1$ for $\psi=1000$. We also computed both lower and higher accuracy metrics for a few values of $\psi$ and observed that the low-lying eigenvalues were relatively insensitive to the change -- for example, at $\psi=100$, a lower accuracy metric with $\sigma=0.1$ gave $\lambda_2=26.1$ while a higher accuracy metric with $\sigma=0.03$ gave $25.8$, agreeing to 1\%.

In Figure~\ref{fig:EigPlots} (left), we show the first 84 eigenvalues and their multiplicities as $\psi$ varies over this range. One sees that the zero-mode $\lambda_1=0$ is always present and that the massive modes appear with multiplicities given by the dimensions of the irreps of the symmetry group~\eqref{eq:irreps}, as expected. The (barely visible) shaded area around each curve corresponds to one standard deviation of the eigenvalue within each set of degenerate eigenvalues.

We see that some eigenmodes become heavier with larger $|\psi|$ while others become lighter, so that the eigenvalues appear to cross as $\psi$ is varied. We can use the multiplicities to track the eigenvalues and find that they do indeed cross rather than simply being too close to distinguish numerically. 

Note that one often encounters no level-crossing or ``avoided crossing'' behavior, e.g.~the von Neumann--Wigner theorem in quantum mechanics or eigenvalue level repulsion in random matrix theory. Indeed, at crossing points, the degeneracy of the eigenvalue is enhanced without an underlying symmetry that would explain this enhanced degeneracy. In the case at hand, level crossing is not a contradiction to this. The degeneracy of the eigenvalues at codimension 0 in complex structure moduli space does have a symmetry origin, and the multiplicities are indeed given by the irreps of the symmetry group $(S_5\times\mathbbm{Z}_2)\ltimes(\mathbbm{Z}_5)^3$. The crossings (and hence the increased degeneracies) occur at codimension one in complex structure moduli space and are hence non-generic. For example around $\psi=6$, $\lambda_2$ and $\lambda_3$ cross, around $\psi=8$  the eigenvalues $\lambda_5$ and $\lambda_6$ cross, etc. In particular, the latter lead to a 25-fold degeneracy at the crossing point, which is not the dimension of an irrep of the symmetry group. Moreover, nothing special seems to be happening (either on the CY or in the complex structure moduli space) at the crossing values for $\psi$, and we are hence led to believe that these enhancements are accidental. Concerning eigenvalue level repulsion in random matrix theory, this then tells us that, while the Laplacian is hermitian, it is far from random for our family of quintics.

We also observe heuristically that the eigenvalues with small degeneracy (i.e.\ irreps with small dimensions) become lighter and the ones with large degeneracy become heavier as $|\psi|$ increases. It would be interesting to understand this better and to predict when level crossing will occur (and for which levels), but we do not have a good understanding of this at the moment.

In Figure~\ref{fig:EigPlots} (right), we show how the first two massive modes, $\lambda_2$ and $\lambda_3$, vary with $d(2,\rho)$, the geodesic distance from $\psi=2$. In particular, we see that $\lambda_3$ falls exponentially with the distance, with the line of best fit given by 
\begin{align}
56.4 \,e^{-(0.906\pm0.034) \,d(2,\rho)}\,,
\end{align}
where the errors show the 95\% confidence interval. Since $m_\text{KK}\sim \lambda^{1/2}$, this suggests that $\alpha\approx 0.45$. This is somewhat smaller than the value of $4/\sqrt 3$ suggested by the analysis around \eqref{eq:psi}. Note however that our result comes from a direct calculation of the $\mathcal{O}(1)$ coefficient for $2\leq \psi \leq 1000$, whereas \eqref{eq:psi} is accurate only in the large complex structure limit. Interestingly, it almost exactly saturates (while being consistent with) the bound $\alpha=1/\sqrt{6}$ proposed in~\cite{Andriot:2020lea}. Applications of a model with $\alpha=1/\sqrt{12}$ to inflation were discussed in~\cite{Cicoli:2017axo}.

In particular, this means that one can move $1/\alpha\approx2.22$ Planck units before the tower comes down one $e$-fold. Moreover, it is actually the second-lightest eigenmode that comes down (the first eigenmode becomes heavier). This means that, due to crossing, the lightest state in the theory reduces by one $e$-fold from $\mathcal{O}(40)$ to $\mathcal{O}(15)$ for a geodesic distance of 3 Planck units. Thus, in concrete cases, trans-Planckian field excursions with a few Planck distances might be feasible. Saying more would, however, require a better understanding of where and when crossing occurs, and why some eigenmodes become heavier for larger $|\psi|$.

The discrepancy between the fitted value of approximately $\alpha=1/\sqrt{6}$ and the infinite distance prediction $\alpha=4/\sqrt 3$ could have several explanations. First, it could be that the swampland distance conjecture needs to be modified to
\begin{align}
m(p_1)=m(p_0)e^{-f(d(p_0,p_1))}\,,
\end{align}
where $f$ is some (non-linear) function that asymptotes to $f(d)=\alpha d$ for $d\to\infty$. While a linear fit of $d$ to $\ln(\lambda)$ certainly fits the data well, there can be seen hints of non-linear behavior in Figure~\ref{fig:EigPlots}. A second possibility is that the asymptotically lightest state has a slope $\alpha=4/\sqrt 3$, but corresponds to a much higher eigenmode $\lambda_n$, which crosses the other levels and becomes the lightest state only at large $\psi$.

\section{Conclusions and Outlook}
\label{sec:Conclusions}
We have used numerical methods to compute the Calabi--Yau metric, the moduli space metric and geodesics, and the spectrum of massive KK modes for the one-parameter family of quintics with varying complex structure $\psi$. From this, we inferred the moduli dependence of the masses of the KK tower and, combining this with the geodesic distance as a function of $\psi$, we found that states become light exponentially, with a coefficient that is of order one, in agreement with the swampland distance conjecture.

There are a number of directions for future work. We have focused on the behavior of the scalar spectrum, however one can repeat this analysis for $(p,q)$-forms. In addition, we considered only a one-parameter family of quintics, however similar conclusions should hold for any of the 101 complex structure moduli. Extending our analysis to this more general case would again require computing the geodesic distance, which could be obtained using the results for moduli space K\"ahler potentials in \cite{Aleshkin:2017oak,Aleshkin:2017fuz,Aleshkin:2018jql}. One could also consider the much larger classes of complete intersection and quotient Calabi--Yau spaces.

We observed level crossing in the eigenvalue spectrum and a qualitatively different behavior of eigenvalues with small or large degeneracies under the symmetry group of the CY. It would be interesting to analyze this further. Since the non-zero eigenvalues encode information about the metric that seemingly cannot be captured by either algebraic geometry or topological data, it is unclear to the authors if this question has an analytic answer, and so a numerical approach seems essential.

Finally, we want to comment on the spectrum at large complex structure. The Strominger--Yau--Zaslow conjecture~\cite{Strominger:1996it} states that any Calabi--Yau threefold is fibered by a special Lagrangian three-torus over a (rational homology) three-sphere. In the large-complex structure limit the three-torus fiber shrinks to zero size, and so the spectrum of the Laplace operator on the threefold should degenerate to the spectrum of the homology sphere. Since the family of quintics we are considering are hypersurfaces in $\mathbbm{P}^4$, they are simply connected and so the base of the fibration must also be simply connected, implying that the base is an honest three-sphere. In principle, one should be able to compute the spectrum restricted to this three-sphere and compare it with the spectrum on the quintic in the large-$\psi$ limit.\footnote{It is somewhat amusing to observe that the first massive mode for large $\psi$ has multiplicity $4$, which is the same as the first massive mode of the round metric on the three-sphere.} One should also be able to see the degeneration of the three-torus using the numerical metric. We hope to come back to this in a future work.

\begin{acknowledgments}
We thank Jim Halverson, Seung-Joo Lee, and Andre Lukas for valuable discussions. The work of AA is supported by the European Union’s Horizon 2020 research and innovation programme under the Marie Sk\l{}odowska-Curie grant agreement No.~838776.
\end{acknowledgments}

\appendix

\section{Einstein frame masses}
\label{app:Einstein}
The SDC is a statement about the behavior of mass scales measured in the Einstein frame. In particular, the scaling of the KK mode masses in \eqref{eq:KKScale} as a function of the K\"ahler modulus is a consequence of some simple facts. Taking the ten-dimensional metric on $M_4\times X$ to be
\begin{equation}
	g_{10} = V^{-1} g_4 + g,
\end{equation}
where $V$ is the volume of $X$ measured by $g$, the four-dimensional metric $g_4$ is automatically in Einstein frame. Given a scalar field $\Phi$ in ten dimensions that satisfies the massless Klein--Gordon equation, we can expand it as $\Phi =  \beta \otimes \phi$ where $\beta$ and $\phi$ depend only on the four- and six-dimensional coordinates respectively. The Klein--Gordon equation can then be written as
\begin{equation}
	\Box \Phi = V\, \Box_4 \beta \otimes \phi + \beta \otimes \Delta \phi,
\end{equation}
where $\Box_4$ is defined by $g_4$ and $\Delta$ is the Laplacian defined by $g$. The mass of the scalar mode $\beta$ in Einstein frame is then simply $m^2 \sim V^{-1} \lambda$, where $\lambda$ is the corresponding eigenvalue of $\Delta$. However, the eigenvalues also scale with the volume of $X$ as $\lambda\sim V^{-1/3}$, so that the squared masses actually scale as $V^{-4/3}$. Combining this with the K\"ahler modulus dependence of the volume, $V\sim r^3$, we recover the $r^{-2}$ scaling given in \eqref{eq:KKScale}.

For the complex structure, things are simpler as the volume factor that takes us to Einstein frame does not depend on $\psi$. The mass scale of the scalar modes then goes as $m^2 \sim \lambda$, where the dependence of $\lambda$ on $\psi$ is what we have explored in the main text.

\section{Symmetries of the family of quintics}
\label{app:Symmetries}
The one-parameter family of quintics~\eqref{eq:quintic} still has many symmetries: it is invariant under permutations of the homogeneous ambient space coordinates, under complex conjugation, and  under multiplying certain combinations of homogeneous coordinates by powers of fifth roots of unity $\xi=e^{2\pi i/5}$. In more detail, we take the generators of the permutation group $S_5$ to be a transposition $t$ and a cyclic permutation $s$,
\begin{align}
\begin{split}
(z_0,z_1,z_2,z_3,z_4)\xrightarrow{t}(z_1,z_0,z_2,z_3,z_4)\,,\\
(z_0,z_1,z_2,z_3,z_4)\xrightarrow{s}(z_1,z_2,z_3,z_4,z_0)\,.
\end{split}
\end{align}
Complex conjugation is just a $\mathbbm{Z}_2$ and we call its generator $c$. The three $\mathbbm{Z}_5$ factors act as
\begin{align}
\begin{split}
(z_0,z_1,z_2,z_3,z_4)\xrightarrow{\mathbbm{Z}_5^{(1)}} (\xi	z_0, \xi^{-1} z_1,z_2, z_3,z_4)\,,\\
(z_0,z_1,z_2,z_3,z_4)\xrightarrow{\mathbbm{Z}_5^{(2)}} (z_0, \xi z_1, \xi^{-1} z_2, z_3,z_4)\,,\\
(z_0,z_1,z_2,z_3,z_4)\xrightarrow{\mathbbm{Z}_5^{(3)}} (z_0, z_1, \xi z_2, \xi^{-1} 	z_3,z_4)\,,
\end{split}
\end{align}
and we call the generators $g_i$, $i=1,2,3$.
Note that the putatively existing ``fourth'' $\mathbbm{Z}_5$ (with generator $g_4$) is not independent, since we also have the projective ambient space symmetry
\begin{align}
(z_0,z_1,z_2,z_3,z_4)\to \lambda (z_0,z_1,z_2,z_3,z_4)\,.
\end{align}
Hence, by choosing $\lambda=\xi$ and calling the generator $\mu$, we find that
\begin{align}
g_4^{-1}=\mu g_3^2 g_2^3 g_1^4\,.
\end{align}
Also note that permutation and complex conjugation commute, the individual $\mathbbm{Z}_5^{(i)}$ phases commute among each other, but permutation and conjugation does not commute with the phases. Hence, the symmetry group is the semi-direct product  $(S_5\times\mathbbm{Z}_2)\ltimes\mathbbm{Z}_5^3$. In order to specify the semi-direct product, we need to specify the twisting, i.e.\ how elements $e\in\mathbbm{Z}_5^3$ change under $g^{-1}\circ e_i\circ g$, where $g$ is an element of $S_5\times\mathbbm{Z}_2$. We hence have to specify the action induced by the generators $c,t,s$ on the generators $g_{1,2,3}$ modulo $\mu$. 

Since complex conjugation will send $\xi\to\xi^{*}=\xi^{-1}$, we find that the induced morphism of $c$ on $\mathbbm{Z}_5^3$ is simply given by inverting the generators,
\begin{align}
(g_1,g_2,g_3)\xrightarrow{c^{-1}\circ e\circ c}(g_1^{-1},g_2^{-1},g_3^{-1})\,.
\end{align}
For the transposition, we compute the action of $t$ on any element $e\in\mathbbm{Z}_5^3$. The element $e$ is labeled by three integers $(n_1,n_2,n_3)\in\mathbbm{Z}_5^3$ that specifies the action
\begin{align*}
\begin{array}{c}
\quad(z_0,z_1,z_2,z_3,z_4)\\[4pt]
\phantom{\mathbbm{Z}_5^3}~~\downarrow \mathbbm{Z}_5^3\\[4pt]
\quad (\xi^{n_1}z_0,\xi^{-n_1+n_2}z_1,\xi^{-n_2+n_3}z_2,\xi^{-n_3}z_3,z_4)\,.
\end{array}
\end{align*}
Hence,
\begin{align}
\begin{array}{c}
(z_0,z_1,z_2,z_3,z_4)\\[4pt]
\phantom{t^{-1}\circ e\circ t}~~\downarrow t^{-1}\circ e\circ t\\[4pt]
(\xi^{-n_1+n_2} z_0,\xi^{n_1}z_1,\xi^{-n_2+n_3}z_2,\xi^{-n_3}z_3,z_4)\,,
\end{array}
\end{align}
such that the transposition acts on the generators $g_i$ as
\begin{align}
(g_1,g_2,g_3)\xrightarrow{t^{-1}\circ e\circ t}(g_2^{-1} g_1, g_2, g_3) \,.
\end{align}
Analogously we find for the action of $c$
\begin{align}
(g_1,g_2,g_3)\xrightarrow{c^{-1}\circ e\circ c}(g_3, g_1g_3^2, g_2g_3^3)
\end{align}
where we also used $\mu$ to remove the action of $c^{-1} \circ e \circ c$ on $z_4$ as well as $5n_i\equiv0$.

The resulting group is of order $2\times 5!\times 5^3=30{,}000$. Now we need to compute the irreps and their dimensions. Note that there can be $n_d>1$ irreps of the same dimension $d$. Given the size of the group, we do this by computing the irreducible characters and their degrees in GAP~\cite{GAP4} via its SAGE interface~\cite{sagemath}. We find a total of 40 irreps with dimensions 1 to 60:
\begin{align}
\label{eq:irreps}
\begin{tabular}{|c|c@{~~~}c@{~~~}c@{~~~}c@{~~~}c@{~~~}c@{~~~}c@{~~~}c@{~~~}c@{~~~}c|}
\hline
$d$&1&4&5&6&20&24&30&40&48&60\\
\hline
$n_d$&4&4&4&2&8&2&8&4&2&2\\
\hline
\end{tabular}
\end{align}

\bibliography{bibliography}

\end{document}